\documentclass[letter, 12pt]{article}

\usepackage[margin=1in]{geometry}
\usepackage[numbers,sort&compress]{natbib} 
\usepackage{authblk} 
\usepackage{color}
\usepackage{amsmath}
\usepackage{amssymb} 
\usepackage{graphicx} 
\usepackage{fancyhdr} 
\usepackage{amsfonts}
\usepackage{tabularx} 
\usepackage{multirow}
\usepackage{slashed}
\usepackage[
singlelinecheck=false 
]{caption}

\usepackage[utf8]{inputenc}

\begin{document}

\renewcommand\headrule{} 

\title{\huge \bf{A Pati-Salam SUSY GUT with Yukawa Unification}}
\author{Zijie Poh$^1$\footnote{poh.7@osu.edu}}
\author{Stuart Raby$^1$\footnote{raby.1@osu.edu}}
\author{Zi-zhi Wang$^2$\footnote{1300011456@pku.edu.cn}}
\affil{$^1$\emph{Department of Physics}\\\emph{The Ohio State University}\\\emph{191 W.~Woodruff Ave, Columbus, OH 43210, USA}}
\affil{$^2$\emph{Department of Physics and State Key Laboratory of Nuclear Physics and Technology}\\\emph{Peking University}\\\emph{5 Yiheyuan Rd, Beijing 100871, P.~R.~China}}

\maketitle
\thispagestyle{fancy}
\pagenumbering{gobble} 

\begin{abstract}\normalsize\parindent 0pt\parskip 5pt
  Previous studies of a three family Yukawa unified supersymmetric grand unified theory (SUSY GUT) with SO(10) or Pati-Salam (PS) gauge symmetry proposed by Raby and students show that this model is able to fit low energy and inflation observables.  However, the fit to low energy observables is not great especially for $\sin2\beta$, and up and down quark masses.  In this paper, we show that by choosing PS as the gauge group and modifying the Yukawa sector, the low energy fit improves significantly while other qualities of the model are maintained.  In particular, the lightest SUSY particle is the neutralino with mass of order 300 - 500 GeV, the lightest stop and sbottom have mass of order 3 - 5 TeV and the CP odd Higgs mass is of order 5 - 6 TeV, so we are in the decoupling limit for the light Higgs.   In addition, we reinterpret gluino simplified model analyses by the ATLAS and CMS collaborations and find that the most stringent gluino mass bound for our model is $M_{\tilde g}\sim1.9\,\text{TeV}$.  The current best fit point, consistent with this bound, with gluino mass $M_{\tilde g}=1.9\,\text{TeV}$ has $\chi^2/\text{dof}\approx1.12$, compared to the best fit point of the previous model with $\chi^2/\text{dof}=1.90$.  We find that predictions for the electric dipole moment of the electron, the branching ratio $BR(\mu \rightarrow e \gamma)$ and the CP violating angle in the lepton sector, $\sin\delta$, are affected significantly as compared to previous results.   In summary, we are unable to rule out this model or place an upper bound on gluino mass accessible by this run of the LHC because the $\chi^2/\text{dof}$ of this model is well below $2\sigma$ even for a gluino mass as high as 2.7~TeV.  On the bright side, this means that this model is still viable and we might find low energy SUSY particles in the near future.

\end{abstract}

\pagenumbering{arabic} 
\section{Introduction}
Supersymmetric grand unified theories (SUSY GUTs) are highly constrained, yet very natural extensions of the Standard Model (SM).  To maintain the possibility of deriving a SUSY GUT from a more fundamental theory such as heterotic string theory, we are interested in SUSY GUTs without large GUT representations.  In particular, this paper discusses a complete three family Yukawa unified SUSY GUT model with either SO(10) or Pati-Salam (PS) $\text{SU}(4)_C\times\text{SU}(2)_L\times\text{SU}(2)_R$ gauge symmetry and a $D_3\times[\text U(1)\times\mathbb Z_2\times\mathbb Z_3]$ family symmetry\footnote{The PS model that we consider in this paper is a GUT because it can be obtained from orbifolding a higher dimensional GUT~\cite{Kobayashi:2004ud,Kobayashi:2004ya}.}.  Previous analyses showed that this model fits reasonably well to low energy observables such as gauge couplings, gauge boson masses, fermion masses, Cabibbo–Kobayashi–Maskawa (CKM) matrix elements, neutrino mass differences and mixing angles~\cite{Dermisek:2005ij,Dermisek:2006dc,Anandakrishnan:2012tj,Poh:2015wta}.  In addition, by including an inflation sector to the PS model, this model can fit the tensor-to-scalar ratio, the scalar spectral index, and the scalar power spectrum~\cite{Bryant:2016tzg}.

Despite the success of this model, as the lower bound on the gluino mass increases, the best chi-squared $\chi^2$ fit of this model to low energy observables is forced to have a chi-squared per degree of freedom $\chi^2/\text{dof}=1.90$~\cite{Poh:2015wta}.  The main contributions to such a large $\chi^2/\text{dof}$ are the poor fits (with pull $\gtrsim3$) to $\sin2\beta$, and the up and down quark masses, $m_u$, $m_d$.   Another major crisis for the model is that supersymmetric partners have not been observed and the most stringent gluino mass bound of a simplified model is $M_{\tilde g}\sim1.9\,\text{TeV}$~\cite{ATLAS:2016uzr}.  This leads to the question of what is the gluino mass bound for this model?  Is this model already ruled out by this bound?  If not, can this model be tested from this run of the LHC?  To answer these questions, recent ATLAS and CMS analyses in signal regions with high jet and $b$-jet multiplicity, leptons, and missing transverse momentum are reinterpreted based on this model.   We show that the data requires that our gluino mass is greater than 1.9 TeV.  In addition, we find that by modifying the Yukawa sector of the model, the low energy fits improve significantly to $\chi^2/\text{dof}=1.12$ for $M_{\tilde g} = 1.9$ TeV.

The rest of the paper is organized as follows.  In Sec.~\ref{sec:ch2.model}, the superpotential of this model along with the Yukawa matrices are given.  We show explicitly the differences between the Yukawa sector of this model and that in previous analyses, and provide some insight on why the modification to the superpotential can improve the low energy fit.  The procedure and the results of a global $\chi^2$ analysis of fitting low energy observables are presented in Sec.~\ref{sec:ch3.chi2}.  Our analysis shows that a huge region in the scalar mass and gluino mass parameter space is $<1.2\sigma$.  Hence, this model is not ruled out by the data.  In Sec.~\ref{sec:ch4.bounds}, we reinterpret the ATLAS and CMS gluino simplified model analyses to obtain the current gluino mass bounds of this model.  The gluino mass bound for our model is $M_{\tilde g}\sim1.9\,\text{TeV}$ which also provides the best fit point (see Fig.~\ref{fig:contour}).  Finally, the discovery prospects and predictions of this model are discussed in Sec.~\ref{sec:ch5.predictions} and we conclude in Sec.~\ref{sec:ch6.conclusion}.

\section{Model}
\label{sec:ch2.model}
The complete three family Yukawa unified SUSY GUT that some of the present authors studied extensively has a SO(10) gauge symmetry with a $D_3\times[\text U(1)\times\mathbb{Z}_2\times\mathbb{Z}_3]$ family symmetry~\cite{Dermisek:2005ij,Dermisek:2006dc,Anandakrishnan:2012tj,Anandakrishnan:2014nea,Poh:2015wta}.  The Yukawa sector superpotential of this model is
\begin{align}
  W = \lambda\ \mathbf{16}_3\ \mathbf{10}\ \mathbf{16}_3 +
      \mathbf{16}_a\ \mathbf{10}\ \chi_a +
      \bar\chi_a\left(
        M_\chi\ \mathbf{\chi}_a +
        \mathbf{45}\ \frac{\phi_a}{\hat M}\ \mathbf{16}_3 +
        \mathbf{45}\ \frac{\tilde\phi_a}{\hat M}\ \mathbf{16}_a +
        \mathbf A\ \mathbf{16}_a\right) \,,
\end{align}
where $\mathbf{16}_i$ is the spinor representation of SO(10), which contains a family of fermions and their supersymmetric partners, and $i=1,2,3$ is the family index.  $\mathbf{16}_3$ is a singlet under $D_3$ symmetry, while $\mathbf{16}_a,a=1,2$ are doublets under $D_3$ symmetry.  $\mathbf{10}$ is the 10 dimensional representation of SO(10), which contains a pair of Higgs doublets.  $\mathbf{45}$ is the adjoint representation of SO(10) that is assumed to obtain vacuum expectation value (VEV) in the $\mathbf B-\mathbf L$ direction.  $\chi_a$ and $\bar\chi_a$ for $a=1,2$ are Froggatt-Neilson states~\cite{Froggatt:1978nt} and are doublets under $D_3$ symmetry.  $\hat M$ is trivial under all groups while $M_\chi=M_0(1+\alpha\mathbf X+\beta\mathbf Y)$, where $\mathbf X$ and $\mathbf Y$ are generators of $SO(10)$, and $\alpha$ and $\beta$ are some constant.  $\mathbf A$ is a SO(10) singlet ``flavon'' field and a non-trivial singlet under $D_3$ symmetry.  Finally, $\phi_a$ and $\tilde\phi_a$ are SO(10) singlet ``flavon'' fields, which are assumed to obtain VEVs of the form
\begin{align}
  \langle\phi_a\rangle
  = \begin{pmatrix}
         \phi_a
      \\ \phi_b
    \end{pmatrix}
  \,,\quad
  \langle\tilde\phi_a\rangle
  = \begin{pmatrix}
         0
      \\ \tilde\phi_2
    \end{pmatrix} \,.
\end{align}

After integrating out the Froggatt-Neilson states, $\chi_a$ and $\bar\chi_a$, and defining
\begin{align}
  G_{x,y} = \frac{M_0}{M_\chi} = \frac{1}{1+\alpha x+\beta y} \,,
\end{align}
along with
\begin{align}
  G^\pm_{x_1,y_1;x_2,y_2} = G_{x_1,y_1}\pm G_{x_2,y_2} \,,
\end{align}
where $x$ and $y$ are the eigenvalues of $\mathbf X$ and $\mathbf Y$, we obtain the following Yukawa matrices\footnote{In previous analyses, we took the limit of $\alpha\gg\beta$ in $G_{x,y}$~\cite{Anandakrishnan:2012tj,Poh:2015wta}.  This limit simplifies the interpretation of the Yukawa matrices but does not reduce the number of input parameters.  Since previous attempts in fitting low energy observables do not produce a good fit, we have decided to include the full $G_{x,y}$ in this analysis.}
\begin{align}
    Y_u &=
    \begin{pmatrix}
        0 &
        \epsilon'G^-_{1,-\frac{4}{3};1,\frac{1}{3}} &
        -\epsilon\xi G_{1,-\frac{4}{3}}
      \\-\epsilon'G^-_{1,-\frac{4}{3};1,\frac{1}{3}} &
        \tilde\epsilon G^-_{1,-\frac{4}{3};1,\frac{1}{3}} &
        -\epsilon G_{1,-\frac{4}{3}}
      \\\epsilon\xi G_{1,\frac{1}{3}} &
        \epsilon G_{1,\frac{1}{3}} &
        \lambda
    \end{pmatrix} \,,
  \\Y_d &=
    \begin{pmatrix}
        0 &
        \epsilon'G^-_{-3,\frac{2}{3};1,\frac{1}{3}} &
        -\epsilon\xi G_{-3,\frac{2}{3}}
      \\-\epsilon'G^-_{-3,\frac{2}{3};1,\frac{1}{3}} &
        \tilde\epsilon G^-_{-3,\frac{2}{3};1,\frac{1}{3}} &
        -\epsilon G_{-3,\frac{2}{3}}
      \\\epsilon\xi G_{1,\frac{1}{3}} &
        \epsilon G_{1,\frac{1}{3}} &
        \lambda
    \end{pmatrix} \,,
  \\Y_e &=
    \begin{pmatrix}
        0 &
        -\epsilon'G^-_{-3,-1;1,2} &
        3\epsilon\xi G_{1,2}
      \\\epsilon'G^-_{-3,-1;1,2} &
        3\tilde\epsilon G^-_{-3,-1;1,2} &
        3\epsilon G_{1,2}
      \\-3\epsilon\xi G_{-3,-1} &
        -3\epsilon G_{-3,-1} &
        \lambda
    \end{pmatrix} \,,
  \\Y_\nu &=
    \begin{pmatrix}
        0 &
        -\epsilon'G^-_{-3,-1;5,0} &
        3\epsilon\xi G_{5,0}
      \\\epsilon'G^-_{-3,-1;5,0} &
        3\tilde\epsilon G^-_{-3,-1;5,0} &
        3\epsilon G_{5,0}
      \\-3\epsilon\xi G_{-3,-1} &
        -3\epsilon G_{-3,-1} &
        \lambda
    \end{pmatrix} \,,
\end{align}
where
\begin{align}
    \epsilon &= -\frac{1}{6}\frac{M_G}{M_0}\frac{\phi_1}{\hat M} \,,
  \\\tilde\epsilon &= +\frac{1}{6}\frac{M_G}{M_0}\frac{\tilde\phi_2}{\hat M} \,,
  \\\epsilon' &= -\frac{1}{2}\frac{A}{M_0} \,,
  \\\xi &= \frac{\phi_2}{\phi_1} \,.
\end{align}
Of these parameters, $\epsilon',\xi,\alpha$ and $\beta$ are complex while $\lambda,\epsilon$ and $\tilde\epsilon$ are real.  Instead of writing the superpotential in SO(10) notation, we can rewrite it using PS fields:
\begin{align}
  \begin{aligned}
    W_{\text{PS}}
      =& \lambda\mathcal Q_3\ \mathcal H\ \mathcal Q_3^c +
         \mathcal Q_a\ \mathcal H\ F_a^c +
         F_a\ \mathcal H\ \mathcal Q_a^c
    \\&+ \bar F_a^c\left(
           M_F\ F_a^c +
           \mathbf{15}\ \frac{\phi_a}{\hat M}\ \mathcal Q_3^c +
           \mathbf{15}\ \frac{\tilde\phi_a}{\hat M}\ \mathcal Q_a^c +
           \mathbf A\ \mathcal Q_a^c\right)
    \\&+ \bar F_a\left(
           M_F\ F_a +
           \mathbf{15}\ \frac{\phi_a}{\hat M}\ \mathcal Q_3 +
           \mathbf{15}\ \frac{\tilde\phi_a}{\hat M}\ \mathcal Q_a +
           \mathbf A\ \mathcal Q_a\right) \,,
  \end{aligned}
\end{align}
where $\{Q_i,F_a\}=(4,2,1),\{Q_i^c,F_a^c\}=(\bar{4},1,\bar{2})$ and $\mathcal H=(1,2,\bar 2)$ under PS symmetry.  $\mathbf{15}$ is the adjoint representation of $\text{SU}(4)_c$ that is assumed to obtain VEV in the $\textbf B-\textbf L$ direction.  In addition, $\bar F_a$ and $\bar F_a^c$ are the conjugate of $F_a$ and $F_a^c$, and $M_F=M_\chi$.  By requiring a PS instead of a SO(10) gauge symmetry, we have more freedom in adding new terms to the superpotential.

In previous global $\chi^2$ analyses, $\sin2\beta$ is too small while $m_u$ and $m_d$ are too large~\cite{Anandakrishnan:2012tj,Poh:2015wta}.  In this analysis, we find that changing $\epsilon'$ to a complex parameter and $\tilde\epsilon$ to a real parameter produces a much better fit for $\sin2\beta$ (see Sec.~3)\footnote{In previous analyses, $\tilde\epsilon$ is a complex parameter and $\epsilon'$ is a real parameter.}.  To improve the fit of $m_u$ and $m_d$, we introduce the following terms to the superpotential\footnote{By adding only the $\Theta'$ terms, we are able to fit $m_u$ and modestly improve the fit of $m_d$.  Having both the $\Theta'$ and $\tilde\Theta$ terms significantly improves the fit of both $m_u$ and $m_d$.}:
\begin{align}
  \bar F_a^c\ \Theta'\ \mathcal Q_a^c +
  \bar F_a\ \Theta'\ \mathcal Q_a +
  \bar F_a^c\ \frac{\tilde\Theta_a}{\hat M}\ \mathcal Q_a^c -
  \bar F_a\ \frac{\tilde\Theta_a}{\hat M}\ \mathcal Q_a \,,
\end{align}
where $\Theta'$ transforms as a trivial singlet and $\tilde\Theta_a,a=1,2$ transforms as doublets under $D_3$ symmetry.  In addition, we assume that $\tilde\Theta_a$ obtains a VEV of the form
\begin{align}
  \langle\tilde\Theta_a\rangle
  = \begin{pmatrix}
      \tilde\Theta_1 \\ 0
    \end{pmatrix} \,,
\end{align}
and both $\Theta'$ and $\tilde\Theta_1$ are real parameters.  With these terms, the Yukawa sector superpotential of this model is
\begin{align}
  \begin{aligned}
    W_{\text{PS}}
      =& \lambda\mathcal Q_3\ \mathcal H\ \mathcal Q_3^c +
         \mathcal Q_a\ \mathcal H\ F_a^c +
         F_a\ \mathcal H\ \mathcal Q_a^c
    \\&+ \bar F_a^c\left(
           M_F\ F_a^c +
           \mathbf{15}\ \frac{\phi_a}{\hat M}\ \mathcal Q_3^c +
           \mathbf{15}\ \frac{\tilde\phi_a}{\hat M}\ \mathcal Q_a^c +
           \mathbf A\ \mathcal Q_a^c +
           \Theta'\ \mathcal Q_a^c +
           \frac{\tilde\Theta_a}{\hat M}\ \mathcal Q_a^c\right)
    \\&+ \bar F_a\left(
           M_F\ F_a +
           \mathbf{15}\ \frac{\phi_a}{\hat M}\ \mathcal Q_3 +
           \mathbf{15}\ \frac{\tilde\phi_a}{\hat M}\ \mathcal Q_a +
           \mathbf A\ \mathcal Q_a +
           \Theta'\ \mathcal Q_a -
           \frac{\tilde\Theta_a}{\hat M}\ \mathcal Q_a\right) \,.
  \end{aligned}
\end{align}
Notice that we want to have a PS gauge symmetry because of the last term in the last two lines of the above equation.  With SO(10) gauge symmetry, we are unable to make these two terms to have opposite sign without introducing a VEV in the $\mathbf B-\mathbf L$ direction.  We find that when we introduce such a VEV, we are able to fit the electron mass, but both $m_u$ and $m_d$ are too large as in our previous analysis.

With the new terms in the Yukawa sector, the Yukawa matrices of this model become
\begin{align}
    Y_u &=
    \begin{pmatrix}
        -\tilde\theta G^-_{1,-\frac{4}{3};1,\frac{1}{3}} &
        \epsilon'G^-_{1,-\frac{4}{3};1,\frac{1}{3}}+\theta'G^+_{1,-\frac{4}{3};1,\frac{1}{3}} &
        -\epsilon\xi G_{1,-\frac{4}{3}}
      \\-\epsilon'G^-_{1,-\frac{4}{3};1,\frac{1}{3}}+\theta'G^+_{1,-\frac{4}{3};1,\frac{1}{3}} &
        \tilde\epsilon G^-_{1,-\frac{4}{3};1,\frac{1}{3}} &
        -\epsilon G_{1,-\frac{4}{3}}
      \\\epsilon\xi G_{1,\frac{1}{3}} &
        \epsilon G_{1,\frac{1}{3}} &
        \lambda
    \end{pmatrix} \,,
  \\Y_d &=
    \begin{pmatrix}
        -\tilde\theta G^-_{-3,\frac{2}{3};1,\frac{1}{3}} &
        \epsilon'G^-_{-3,\frac{2}{3};1,\frac{1}{3}}+\theta'G^+_{-3,\frac{2}{3};1,\frac{1}{3}} &
        -\epsilon\xi G_{-3,\frac{2}{3}}
      \\-\epsilon'G^-_{-3,\frac{2}{3};1,\frac{1}{3}}+\theta'G^+_{-3,\frac{2}{3};1,\frac{1}{3}} &
        \tilde\epsilon G^-_{-3,\frac{2}{3};1,\frac{1}{3}} &
        -\epsilon G_{-3,\frac{2}{3}}
      \\\epsilon\xi G_{1,\frac{1}{3}} &
        \epsilon G_{1,\frac{1}{3}} &
        \lambda
    \end{pmatrix} \,,
  \\Y_e &=
    \begin{pmatrix}
        \tilde\theta G^-_{-3,-1;1,2} &
        -\epsilon'G^-_{-3,-1;1,2}+\theta'G^+_{-3,-1;1,2} &
        3\epsilon\xi G_{1,2}
      \\\epsilon'G^-_{-3,-1;1,2}+\theta'G^+_{-3,-1;1,2} &
        3\tilde\epsilon G^-_{-3,-1;1,2} &
        3\epsilon G_{1,2}
      \\-3\epsilon\xi G_{-3,-1} &
        -3\epsilon G_{-3,-1} &
        \lambda
    \end{pmatrix} \,,
  \\Y_\nu &=
    \begin{pmatrix}
        \tilde\theta G^-_{-3,-1;5,0} &
        -\epsilon'G^-_{-3,-1;5,0}+\theta'G^+_{-3,-1;5,0} &
        3\epsilon\xi G_{5,0}
      \\\epsilon'G^-_{-3,-1;5,0}+\theta'G^+_{-3,-1;5,0} &
        3\tilde\epsilon G^-_{-3,-1;5,0} &
        3\epsilon G_{5,0}
      \\-3\epsilon\xi G_{-3,-1} &
        -3\epsilon G_{-3,-1} &
        \lambda
    \end{pmatrix} \,,
\end{align}
where
\begin{align}
    \theta' &= -\frac{1}{2}\frac{\Theta'}{M_0} \,,
  \\\tilde\theta &= +\frac{1}{2}\frac{M_G}{M_0}\frac{\tilde\Theta_1}{\hat M} \,.
\end{align}

In summary, our model has 26 input parameters (see Tab.~\ref{tab:input}).  The fermion sector has 17 parameters - 13 Yukawa parameters, $\tan\beta$, and 3 right-handed neutrino masses, while the SM has 19 observables - 9 fermion masses, 4 CKM matrix elements, 2 neutrino mass differences, 3 real neutrino mixing angles, and 1 neutrino CP violating phase.  Hence, our model has 2 predictions in the fermion sector.  As a comparison with previous analyses, we have added two new real input parameters in the Yukawa sector.

\begin{table}
  \begin{center}
    \renewcommand{\arraystretch}{1.2}
      \scalebox{0.83}{
        \begin{tabular}{|l||c|c|}
          \hline Sector & Input Parameters & No.
          \\ \hline Gauge & $\alpha_G$, $M_G$, $\epsilon_3$ & 3
          \\ SUSY (GUT scale) & $m_{16}$, $M_{1/2}$, $A_0$, $m_{H_u}$, $m_{H_d}$ & 5
          \\ Yukawa Textures & $\lambda,\epsilon,\tilde\epsilon,\epsilon',\xi,\alpha,\beta,\theta',\tilde\theta,\phi_{\epsilon'},\phi_\xi,\phi_\alpha,\phi_\beta$ & 13
          \\ Neutrino & $M_{R_1}$, $M_{R_2}$, $M_{R_3}$ & 3
          \\ SUSY (EW Scale) & $\tan\beta$, $\mu$ & 2
          \\ \hline Total & & 26
          \\ \hline
        \end{tabular}
      }
    \caption{
    Our model has 26 input parameters.
    }
    \label{tab:input}
  \end{center}
\end{table}

\section{Global Chi-Squared Analysis}
\label{sec:ch3.chi2}
\subsection{Procedure}
The program that performs the renormalization group equation (RGE) running and calculation of low energy observables, \texttt{maton}, is developed in-house by Radovan Derm\'i\v{s}ek.  This program adopts a top-down approach; that is the program starts with input parameters at the GUT scale and uses two-loop RGE along with one-loop threshold corrections to run to the low energy scale where observables are calculated.  The calculated observables are then compared with experimental measurements in a chi-squared equation:
\begin{align}
  \chi^2 = \sum_i\frac{|x_i^\text{th}-x_i^\text{exp}|^2}{\sigma_i^2} \,,
\end{align}
where $x_i^\text{th}$ are the calculated values, $x_i^\text{exp}$ are the measured values, and $\sigma_i^2$ is the sum of the squares of theoretical and experimental uncertainties.  The program then uses the \texttt{Minuit} package~\cite{James:1975dr} to minimize this $\chi^2$ function.  The details of the program can be found in previous analyses~\cite{Anandakrishnan:2012tj,Poh:2015wta}.

In this paper, we fit this model to 51 observables listed in Tab.~\ref{tab:obs}.  As a comparison with previous analyses, we have included $m_u$ and $m_d$ to the list of observables.  In addition, we have updated all experimental values to the latest value available in the Particle Data Group (PDG) and Heavy Flavor Averaging Group~\cite{Olive:2016xmw,Amhis:2016xyh}.  The value of $|V_{cb}|$ and $|V_{ub}|$ used are the average of the inclusive and exclusive values from the PDG with error bars overlapping the inclusive and exclusive error bars.  We also updated the publicly available software, \texttt{superiso} and \texttt{susyflavor}~\cite{Mahmoudi:2008tp,Rosiek:2014sia}.  Every $\chi^2$ minimization of this paper is done by fixing $m_{16}$ and $M_{1/2}$ in a grid of points and then minimizing with respect to the other 24 parameters.  Since we are fitting to 51 observables, for each fixed value of $m_{16}$ and $M_{1/2}$, we have 27 dof.

\begin{table}[!htbp]\footnotesize
  \begin{center}
  \scalebox{0.7}{
    \renewcommand{\arraystretch}{1.0}
    \begin{tabular}{|l|l|r|c|r|}
      \hline
        \multicolumn{1}{|c|}{\textbf{Observable}} &
        \multicolumn{1}{c|}{\textbf{Exp.~Value}} &
        \multicolumn{1}{c|}{\textbf{Ref.}} &
        \multicolumn{1}{c|}{\textbf{Program}} &
        \multicolumn{1}{c|}{\textbf{Th.~Error}}
      \\\hline\hline
        $M_Z$ &
          $91.1876\pm0.0021\,\text{GeV}$ &
            \cite{Olive:2016xmw} &
              Input &
                0.0\%\ \ \
      \\$M_W$ &
          $80.385\pm0.015\,\text{GeV}$ &
            \cite{Olive:2016xmw} &
              \texttt{maton} &
                0.5\%\ \ \
      \\$\alpha_\text{em}$ &
          $1/137.035999139(31)$ &
            \cite{Olive:2016xmw} &
              \texttt{maton} &
                0.5\%\ \ \
      \\$G_\mu$ &
          $1.1663787(6)\times10^{-5}\,\text{GeV}^{-2}$ &
            \cite{Olive:2016xmw} &
              \texttt{maton} &
                1.0\%\ \ \
      \\$\alpha_3(M_Z)$ &
          $0.1181\pm0.0006$ &
            \cite{Olive:2016xmw} &
              \texttt{maton} &
                0.5\%\ \ \
      \\\hline
        $M_t$ &
          $173.21\pm0.51\pm0.71\,\text{GeV}$ &
            \cite{Olive:2016xmw} &
              \texttt{maton} &
                1.1\%\ \ \
      \\$m_b(m_b)$ &
          $4.185\pm0.035\,\text{GeV}$ &
            \cite{Olive:2016xmw} &
              \texttt{maton} &
                3.0\%\ \ \
      \\$M_\tau$ &
          $1776.86\pm0.12\,\text{MeV}$ &
            \cite{Olive:2016xmw} &
              \texttt{maton} &
                1.1\%\ \ \
      \\\hline
        $m_b-m_c$ &
          $3.45\pm0.05\,\text{GeV}$ &
            \cite{Olive:2016xmw} &
              \texttt{maton} &
                10.8\%\ \ \
      \\$m_c(m_c)$ &
          $1.27\pm0.03\,\text{GeV}$ &
            \cite{Olive:2016xmw} &
              \texttt{maton} &
                1.1\%\ \ \
      \\$m_s(2\,\text{GeV})$ &
          $98\pm6\,\text{MeV}$ &
            \cite{Olive:2016xmw} &
              \texttt{maton} &
                1.1\%\ \ \
      \\$m_s/m_d\,(2\,\text{GeV})$ &
          $19.5\pm2.5$ &
            \cite{Olive:2016xmw} &
              \texttt{maton} &
                0.5\%\ \ \
      \\$Q$ &
          $23\pm2$ &
            \cite{Olive:2016xmw} &
              \texttt{maton} &
                5.0\%\ \ \
      \\$m_u\,(2\ \text{GeV})$ &
          $2.3\pm0.5\,\text{MeV}$ &
            \cite{Olive:2016xmw} &
              \texttt{maton} &
                1.1\%\ \ \
      \\$m_d\,(2\ \text{GeV})$ &
          $4.75\pm0.45\,\text{MeV}$ &
            \cite{Olive:2016xmw} &
              \texttt{maton} &
                1.1\%\ \ \
      \\$M_\mu$ &
          $105.6583745(24)\,\text{MeV}$ &
            \cite{Olive:2016xmw} &
              \texttt{maton} &
                2.1\%\ \ \
      \\$M_e$ &
          $0.5109989461(31)\,\text{MeV}$ &
            \cite{Olive:2016xmw} &
              \texttt{maton} &
                1.1\%\ \ \
      \\\hline
        $|V_{ud}|$ &
          $0.97417\pm0.00021$ &
            \cite{Olive:2016xmw} &
              \texttt{maton} &
                0.5\%\ \ \
      \\$|V_{us}|$ &
          $0.2248\pm0.0006$ &
            \cite{Olive:2016xmw} &
              \texttt{maton} &
                0.5\%\ \ \
      \\$|V_{ub}|$ &
          $(4.13\pm0.60)\times10^{-3}$ &
            \cite{Olive:2016xmw} &
              \texttt{maton} &
                2.1\%\ \ \
      \\$|V_{cd}|$ &
          $0.220\pm0.005$ &
            \cite{Olive:2016xmw} &
              \texttt{maton} &
                0.5\%\ \ \
      \\$|V_{cs}|$ &
          $0.995\pm0.016$ &
            \cite{Olive:2016xmw} &
              \texttt{maton} &
                0.5\%\ \ \
      \\$|V_{cb}|$ &
          $(40.75\pm2.25)\times10^{-3}$ &
            \cite{Olive:2016xmw} &
              \texttt{maton} &
                2.1\%\ \ \
      \\$|V_{td}|$ &
          $(8.2\pm0.6)\times10^{-3}$ &
            \cite{Olive:2016xmw} &
              \texttt{maton} &
                2.1\%\ \ \
      \\$|V_{ts}|$ &
          $(40.0\pm2.7)\times10^{-3}$ &
            \cite{Olive:2016xmw} &
              \texttt{maton} &
                2.1\%\ \ \
      \\$|V_{tb}|$ &
          $1.009\pm0.031$ &
            \cite{Olive:2016xmw} &
              \texttt{maton} &
                0.5\%\ \ \
      \\$\sin2\beta$ &
          $0.691\pm0.017$ &
            \cite{Olive:2016xmw} &
              \texttt{maton} &
                0.5\%\ \ \
      \\$\epsilon_K$ &
          $(2.233\pm0.015)\times10^{-3}$ &
            \cite{Olive:2016xmw} &
              \texttt{susyflavor}\cite{Rosiek:2014sia} &
                10.0\%\ \ \
      \\\hline
        $\Delta m_{B_s}/\Delta m_{B_d}$ &
          $34.8479\pm0.2324$ &
            \cite{Olive:2016xmw} &
              \texttt{susyflavor}\cite{Rosiek:2014sia} &
                20.2\%\ \ \
      \\$\Delta m_{B_d}$ &
          $(3.354\pm0.022)\times 10^{-10}\,\text{MeV}$ &
            \cite{Olive:2016xmw} &
              \texttt{susyflavor}\cite{Rosiek:2014sia} &
                20.0\%\ \ \
      \\\hline
        $\Delta m_{21}^2$ &
          $(7.375\pm0.165)\times10^{-5}\,\text{eV}^2$ &
            \cite{Capozzi:2016rtj} &
              \texttt{maton} &
                5.0\%\ \ \
      \\$\Delta m_{31}^2$ &
          $(2.50\pm0.04)\times10^{-3}\,\text{eV}^2$ &
            \cite{Capozzi:2016rtj} &
              \texttt{maton} &
                5.0\%\ \ \
      \\$\sin^2\theta_{12}$ &
          $0.2975\pm0.0165$ &
            \cite{Capozzi:2016rtj} &
              \texttt{maton} &
                0.5\%\ \ \
      \\$\sin^2\theta_{23}$ &
          $0.4435\pm0.0265$ &
            \cite{Capozzi:2016rtj} &
              \texttt{maton} &
                0.5\%\ \ \
      \\$\sin^2\theta_{13}$ &
          $0.0215\pm0.0010$ &
            \cite{Capozzi:2016rtj} &
              \texttt{maton} &
                0.5\%\ \ \
      \\\hline
        $M_h$ &
          $125.90\pm0.24\,\text{GeV}$ &
            \cite{Olive:2016xmw} &
              \texttt{splitsuspect}\cite{Bernal:2007uv} &
                3.8\%\ \ \
      \\\hline
        $\text{BR}(b\rightarrow s\gamma)$ &
          $(332\pm16)\times10^{-6}$ &
            \cite{Amhis:2016xyh} &
              \texttt{susyflavor}\cite{Rosiek:2014sia} &
                47.3\%\ \ \
      \\$\text{BR}(B_s\rightarrow\mu^+\mu^-)$ &
          $(2.94\pm0.65)\times10^{-9}$ &
            \cite{Amhis:2016xyh} &
              \texttt{susyflavor}\cite{Rosiek:2014sia} &
                22.4\%\ \ \
      \\$\text{BR}(B_d\rightarrow\mu^+\mu^-)$ &
          $(0.40\pm0.15)\times10^{-9}$ &
            \cite{Amhis:2016xyh} &
              \texttt{susyflavor}\cite{Rosiek:2014sia} &
                21.5\%\ \ \
      \\$\text{BR}(B\rightarrow\tau\nu)$ &
          $(106\pm19)\times10^{-6}$ &
            \cite{Amhis:2016xyh} &
              \texttt{susyflavor}\cite{Rosiek:2014sia} &
                50.4\%\ \ \
      \\\hline
        $\text{BR}(B\rightarrow K^*\mu^+\mu^-)_{1\leq q^2\leq6\,\text{GeV}^2}$  &
           $(0.34\pm0.06)\times10^{-7}$ &
            \cite{Aaij:2013iag} &
              \texttt{superiso}\cite{Mahmoudi:2008tp} &
                105.1\%\ \ \
      \\$\text{BR}(B\rightarrow K^*\mu^+\mu^-)_{14.18\leq q^2\leq16\,\text{GeV}^2}$ &
          $(0.56\pm0.10)\times10^{-7}$ &
            \cite{Aaij:2013iag} &
              \texttt{superiso}\cite{Mahmoudi:2008tp} &
                190.0\%\ \ \
      \\$q_0^2(\text{A}_\text{FB}(B\rightarrow K^*\mu^+\mu^-))$ &
          $4.9\pm0.9\,\text{GeV}^2$ &
            \cite{Aaij:2013iag} &
              \texttt{superiso}\cite{Mahmoudi:2008tp} &
                25.3\%\ \ \
      \\$F_L(B\rightarrow K^*\mu^+\mu^-)_{1\leq q^2\leq6\,\text{GeV}^2}$ &
          $0.65\pm0.09$ &
            \cite{Aaij:2013iag} &
              \texttt{superiso}\cite{Mahmoudi:2008tp} &
                45.0\%\ \ \
      \\$F_L(B\rightarrow K^*\mu^+\mu^-)_{14.18\leq q^2\leq16\,\text{GeV}^2}$ &
          $0.33\pm0.09$ &
            \cite{Aaij:2013iag} &
              \texttt{superiso}\cite{Mahmoudi:2008tp} &
                80.0\%\ \ \
      \\$-2P_2=A_T^\text{Re}(B\rightarrow K^*\mu^+\mu^-)_{1\leq q^2\leq6\,\text{GeV}^2}$ &
          $-0.66\pm0.24$ &
            \cite{Aaij:2013iag} &
              \texttt{superiso}\cite{Mahmoudi:2008tp} &
                198.2\%\ \ \
      \\$-2P_2=A_T^\text{Re}(B\rightarrow K^*\mu^+\mu^-)_{14.18\leq q^2\leq16\,\text{GeV}^2}$ &
          $0.50\pm0.03$ &
            \cite{Aaij:2013iag} &
              \texttt{superiso}\cite{Mahmoudi:2008tp} &
                45.0\%\ \ \
      \\$P_4'(B\rightarrow K^*\mu^+\mu^-)_{1\leq q^2\leq 6\,\text{GeV}^2}$ &
          $0.58\pm0.36$ &
            \cite{Aaij:2013qta} &
              \texttt{superiso}\cite{Mahmoudi:2008tp} &
                30.4\%\ \ \  
      \\$P_4'(B\rightarrow K^*\mu^+\mu^-)_{14.18\leq q^2\leq16\,\text{GeV}^2}$ &
          $-0.18\pm0.70$ &
            \cite{Aaij:2013qta} &
              \texttt{superiso}\cite{Mahmoudi:2008tp} &
                35.0\%\ \ \  
      \\$P_5'(B\rightarrow K^*\mu^+\mu^-)_{1\leq q^2\leq6\,\text{GeV}^2}$ &
          $0.21\pm0.21$ &
            \cite{Aaij:2013qta} &
              \texttt{superiso}\cite{Mahmoudi:2008tp} &
                45.9\%\ \ \
      \\$P_5'(B\rightarrow K^*\mu^+\mu^-)_{14.18\leq q^2\leq16\,\text{GeV}^2}$ &
          $-0.79\pm0.27$ &
            \cite{Aaij:2013qta} &
              \texttt{superiso}\cite{Mahmoudi:2008tp} &
                60.0\%\ \ \
      \\\hline
    \end{tabular}
  }
  \end{center}
  \caption{
    51 low-energy observables that are fitted in the global $\chi^2$ analysis.
  }
  \label{tab:obs}
\end{table}

A way to visualize the gluino mass that this model favors is to make a $\chi^2/\text{dof}$ contour plot of the gluino mass as a function of the scalar mass at the GUT scale, $m_{16}$.  To produce this plot, we perform the $\chi^2$ minimization for gluino mass ranges from 1.6~TeV to 2.8~TeV with an increment of 0.2~TeV and scalar mass at the GUT scale ranging from 10~TeV to 30~TeV with an increment of 5~TeV.  We control the gluino mass by selecting the appropriate value of $M_{1/2}$ at the GUT scale.  We then perform a two dimensional cubic spline interpolation on these 30 points to obtain a two dimension surface of $\chi^2/\text{dof}$.  To increase the likelihood that each of these 30 points is at the minimum, we perform minimization repeatedly until the change in $\chi^2$ after 5 repetitions is lower than 0.001.  After the $\chi^2$ value settles down, we make a small shift in the values of input parameters other than $M_{1/2}$ and $m_{16}$ and re-perform the minimization to make sure that the shifted parameters eventually return to the original value.  With this procedure, we are confident that the points that we obtain are at least in a very deep local minimum, if not the global minimum.

\subsection{Results}
\begin{figure}
  \begin{center}
    \includegraphics[width=0.85\textwidth]{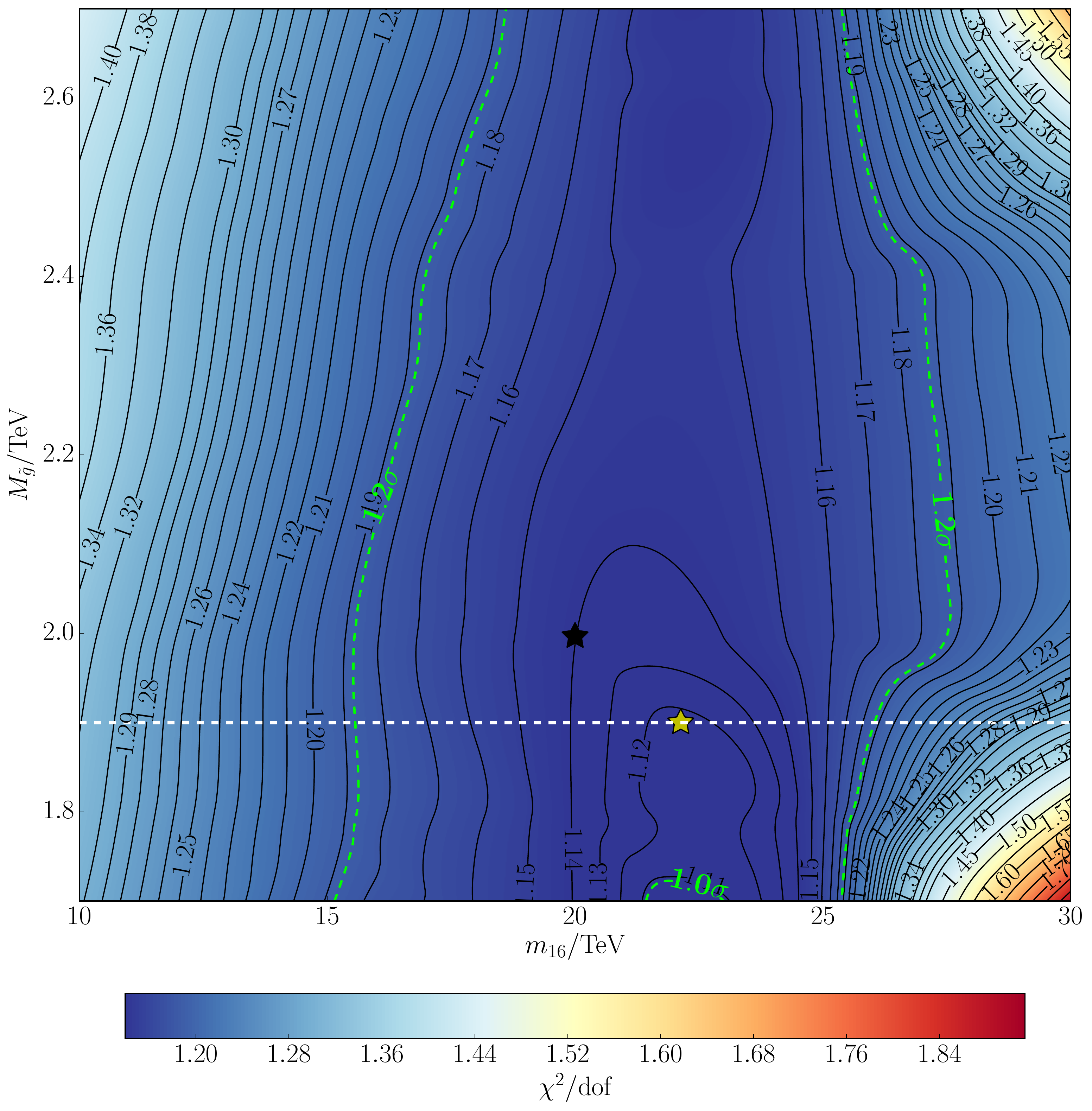}
    \caption{
    $\chi^2/\text{dof}$ contour plot of gluino mass, $M_{\tilde g}$ as a function of scalar mass at GUT scale, $m_{16}$.  The green dotted lines are the 1.0 and 1.2 $\sigma$ bound from the $\chi^2$ analysis with 27 dof.  This plot has 27 dof because $m_{16}$ and $M_{1/2}$ are fixed as the $x$ and $y$-axis. The horizontal white dotted line is the current gluino mass bound of our model, see Sec.~\ref{sec:ch4.bounds}.  The yellow star is the point with the lowest $\chi^2$ for gluino mass above the current bound.  The black star is a benchmark point where its input parameters and low energy fits are shown in the appendix.  Since the global $\chi^2$ minimum is below the lower limit of the plot, our model prefers low gluino mass.  However, this plot also shows that our model is not very sensitive to the gluino mass, because $\chi^2$ increases relatively slowly as the gluino mass increases.
    }
    \label{fig:contour}
  \end{center}
\end{figure}

Fig.~\ref{fig:contour} shows the $\chi^2/\text{dof}$ contour plot with gluino mass ranging from 1.7 TeV to 2.7 TeV and the scalar mass at the GUT scale ranging from 10 TeV to 30 TeV.  The values of $\chi^2/\text{dof}$ ranges from 1.10 to 1.89.  The black contour lines show that our model prefers small gluino mass because the minimum $\chi^2$ value occurs at gluino mass below the lower limit of the plot.  The green dotted lines are the 1.0 and 1.2 $\sigma$ bound.  These lines show that even with $M_{\tilde g}=2.7\,\text{TeV}$, $\chi^2/\text{dof}$ can still be as low as $\approx1.15$, which is well within the $2\sigma$ bound\footnote{In fact, even with gluino mass as high as 3.1 TeV and $m_{16}=25\,\text{TeV}$, $\chi^2/\text{dof}=1.33$, which is also below the $2\sigma$ bound.}.  Since the $\chi^2$ contour lines are very flat in the gluino mass direction, this model is not very sensitive to gluino mass.  Hence, this model cannot be ruled out even if the gluino is not seen during the current run of the LHC.

Also shown in Fig.~\ref{fig:contour} are a horizontal white line, and a black, and a yellow star.  The horizontal white line is the current gluino mass bound obtained by reinterpreting the most recent gluino mass bound from the ATLAS and CMS colaborations (see next section).  The black star is a benchmark point.  The input parameters and the low energy fits of this benchmark point are shown in the appendix.  The yellow star is the point with the minimum $\chi^2$ with gluino mass still allowed by current bound.  Notice that this point is exactly on the current mass bound indicating that even though the $\chi^2/\text{dof}$ is relatively small for large gluino mass, our model still prefers small gluino mass.

\section{Current LHC Bounds}
\label{sec:ch4.bounds}
The typical search for supersymmetry is performed under the assumption of a simplified model, such as T1tttt or Gtt model, in which the gluino decays 100\% of the time to $t\bar{t}\tilde{\chi}_1^0$.  On the other hand, gluinos in this model do not decay via a single channel (see Tab.~\ref{tab:gluino_br} for typical branching ratios of gluino of this model).  Hence, to obtain the current gluino bound for this model, the analyses performed by the ATLAS and CMS collaborations has to be reinterpreted.

\begin{table}[!htbp]\footnotesize
  \begin{center}
    \renewcommand{\arraystretch}{1.0}
    \begin{tabular}{|l|r|r|}
      \hline
      $m_{16}/\text{TeV}$ & 20 & 25 \\
      $M_{\tilde g}/\text{TeV}$ & 1.90 & 1.90 \\\hline\hline
      $g\tilde\chi_1^0$ & 0.000 & 0.000 \\
      $g\tilde\chi_2^0$ & 0.002 & 0.001 \\
      $g\tilde\chi_3^0$ & 0.005 & 0.007 \\
      $g\tilde\chi_4^0$ & 0.002 & 0.004 \\\hline
      $tb\tilde\chi_1^+$ & 0.234 & 0.186 \\
      $tb\tilde\chi_2^+$ & 0.274 & 0.322 \\
      $t\bar t\tilde\chi_1^0$ & 0.019 & 0.023 \\
      $t\bar t\tilde\chi_2^0$ & 0.054 & 0.039 \\
      $t\bar t\tilde\chi_3^0$ & 0.113 & 0.105 \\
      $t\bar t\tilde\chi_4^0$ & 0.097 & 0.106 \\
      $b\bar b\tilde\chi_1^0$ & 0.010 & 0.011 \\
      $b\bar b\tilde\chi_2^0$ & 0.064 & 0.054 \\
      $b\bar b\tilde\chi_3^0$ & 0.082 & 0.082 \\
      $b\bar b\tilde\chi_4^0$ & 0.044 & 0.059 \\\hline
    \end{tabular}
  \end{center}
  \caption{
    Gluino branching ratios of a point with $m_{16}=20\,\text{TeV}$ and another point with $m_{16}=25\,\text{TeV}$.  Both points have $M_{\tilde g}=1.90\,\text{TeV}$.  This table shows that gluino branching ratios of our model is not even close to that of the simplified model.
  }
  \label{tab:gluino_br}
\end{table}

The most stringent gluino mass bound is from \uppercase{atlas-conf-2016-052}, where the gluino mass bound of the Gtt simplified model is around 1.9~TeV with the lightest neutralino mass $m_{\tilde{\chi}_1^0}=200\,\text{GeV}$~\cite{ATLAS:2016uzr}.  \uppercase{atlas-conf-2016-052} considers the signal region with zero or more leptons, $b$-jets and missing transverse momentum.  The most stringent gluino mass bound from the CMS collaboration is from \uppercase{cms-sus-16-014}, where the gluino mass bound of the T1tttt simplified model is 1.75~TeV for $m_{\tilde{\chi}_1^0}=200\,\text{GeV}$~\cite{CMS:2016mwj}.  \uppercase{cms-sus-16-021} considers the signal region with jets and missing transverse momentum.  In this paper, these two analyses are reinterpreted with our model.

In addition, a CMS analysis, \uppercase{cms-sus-16-021}, which considers the signal region of two opposite-sign same-flavor leptons with jets and missing transverse momentum, found a $2.1(1.1)\sigma$ local(global) deviation in the number of observed events compared to the SM background~\cite{CMS:2016zwk}.  Since our model produces signal in this region, we include this analysis in this paper.  To be impartial, we also reinterpret a CMS analysis that consider signal region with same-sign dilepton events, \uppercase{cms-sus-16-020}~\cite{CMS:2016vfu}.

The experimental data, for all the analyses mentioned above, is obtained at the center-of-mass energy $\sqrt{s}=13\,\text{TeV}$.  The integrated luminosity for the ATLAS analysis is $14.8\,\text{fb}^{-1}$, while that of the CMS analyses is $12.9\,\text{fb}^{-1}$.

\subsection{Analysis Procedure}
In a nutshell, the analyses are re-performed by focusing on the 95\% upper limit of the number of events allowed, $N_\text{UL}$, calculated from the SM background and the number of observed events.  $N_\text{UL}$ is the 95\% Bayesian upper limit for a Poisson parameter calculated using a uniform prior.  By focusing on the number of events allowed, we do not need to perform background simulation.  Instead, we only need to simulate events produced by the models in consideration, such as the simplified model and our model.  To validate our analysis, we first ensure that the simplified mass bound obtained from our analysis matches with those from the ATLAS and CMS analyses. Mass bounds are obtained by ruling out masses where the 95\% lower limit on the number of events passing all cuts exceeds $N_\text{UL}$.  Once our analysis is validated, we can re-perform the analysis based on our model to obtain the gluino mass bound of our model.

For analysis of a simplified model, events are simulated by supplying \texttt{\uppercase{pythia} 8.219} with an SLHA file that contains the SUSY spectrum, mixing angles and decay tables~\cite{Sjostrand:2014zea}.  For each mass point, 10,000 events are simulated.  The simulated events are then passed to \texttt{Delphes 3.4.0}, a detector simulator that outputs events as recorded by the detector~\cite{deFavereau:2013fsa}.  The card files of \texttt{Delphes}, which specifies various detector specific parameters such as the triggering and candidate selection requirements, are modified according to the selection criteria of the ATLAS and CMS analyses.  The output of \texttt{Delphes} then goes through a cutflow code that we wrote.  The number of events passing all cuts is then normalized by the ratio of the number of events produced at LHC to the number of simulated events.  The number of events produced at LHC equals the product of the luminosity and the production cross section, which is obtained from the LHC SUSY Cross Section Working Group~\cite{Borschensky:2014cia}.  The normalized number of events passing all cuts is then compared to $N_\text{UL}$ to produce the mass bound.

The analysis of our model is almost identical to that of the simplified model.  The only difference is that the SLHA file of the simplified model is simple and can be written directly by hand, while that of our model is very complicated.  Luckily, \texttt{maton} is also a spectrum generator.  After obtaining the SUSY spectrum along with all the mixing angles and couplings of the model, we use \texttt{SUSY-HIT 1.5a} to calculate the decay tables~\cite{Djouadi:2006bz}.  The output of \texttt{SUSY-HIT} is then used as input to \texttt{PYTHIA} and the procedure of the simplified model analysis outlined in previous paragraph is repeated.

\subsection{Results}
Out of all the signal regions in the four analyses that we studied, the most constrained bound comes from the 0-lepton with large mass splitting signal region in the ATLAS analysis, \uppercase{atlas-conf-2016-52}.  The events in this signal region are required to have $N^\text{signal lepton}=0$, $N^\text{jet}\geq8$, $N_{b-\text{jet}}\geq3, p_T^\text{jet}>30\,\text{GeV}$, $E_T^\text{miss}>400\,\text{GeV}$, $\Delta\phi_\text{min}^{4j}>0.4\,\text{rad}$, $m_\text{T,min}^{b-\text{jets}}>80\,\text{GeV}$, $m_\text{eff}^\text{incl}>2000\,\text{GeV}$ and $M_J^\Sigma>200\,\text{GeV}$.  These parameters are defined in \cite{ATLAS:2016uzr}.  Hence, in this section, we will only show the results of this specific analysis.

\begin{figure}
  \begin{center}
    \includegraphics[width=0.7\textwidth]{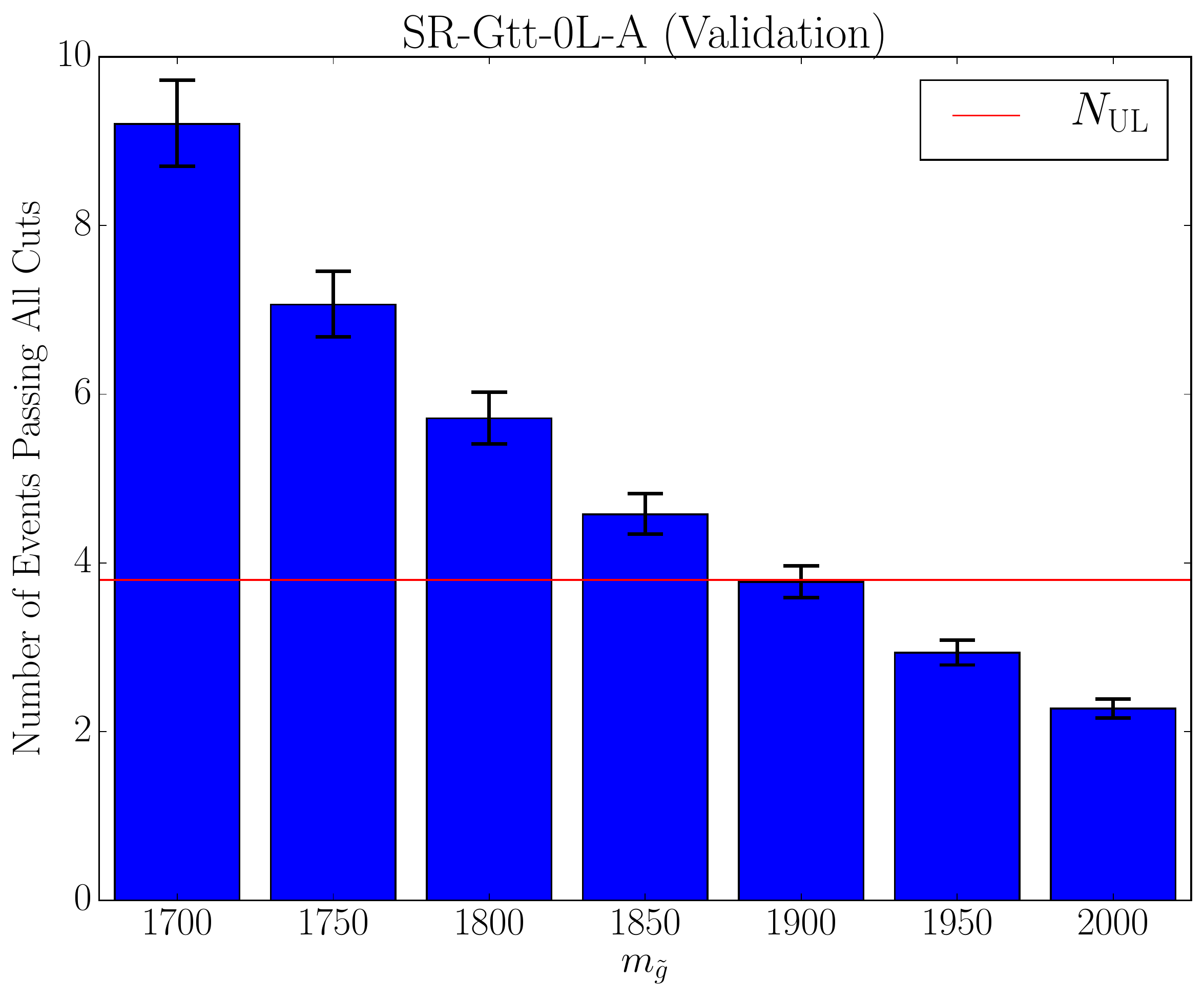}
    \caption{
    The validation plot for Gtt simplified model with $m_{\tilde\chi_1^0}=200\,\text{GeV}$ in the 0-lepton with large mass splitting signal region of the ATLAS analysis~\cite{ATLAS:2016uzr}.  The vertical blue bars show the number of events passing all cuts while the horizontal red line is the 95\% upper limit of the number of events allowed.  The gluino mass bound obtained from this plot, $M_{\tilde g}\sim1.875\,\text{TeV}$, is in agreement with the gluino mass bound from the ATLAS collaboration, $M_{\tilde g}\sim1.9\,\text{TeV}$.
    }
    \label{fig:gtt0la.valid}
  \end{center}
\end{figure}

Fig.~\ref{fig:gtt0la.valid} is the validation plot from our analysis of the Gtt simplified model with $m_{\tilde\chi_1^0}=200\,\text{GeV}$.  The red horizontal line is the 95\% upper limit of the number of events allowed, $N_\text{UL}=3.8$.  The vertical blue bars are the normalized number of events passing all cuts.  The error bars represent the 95\% upper and lower limits of the number of events passing all cuts.  These limits are derived from the uncertainties in the gluino production cross section and the counting experiment.  The size of the error bars shrink as the gluino mass increases because the number of simulated events stays constant but the gluino production cross section decreases.  Fig.~\ref{fig:gtt0la.valid} shows that the gluino mass bound from our analysis is $M_{\tilde g}\sim1.875\,\text{TeV}$, which is well within 20\% of the gluino mass bounds from the ATLAS analysis, $M_{\tilde g}\sim1.9\,\text{TeV}$.  This is the expected precision because we do not have the state of the art analysis tools available to the ATLAS collaboration, such as the detector simulator.  From this, we conclude that our analysis is in agreement with the ATLAS analysis.
\begin{figure}
  \begin{center}
    \includegraphics[width=0.7\textwidth]{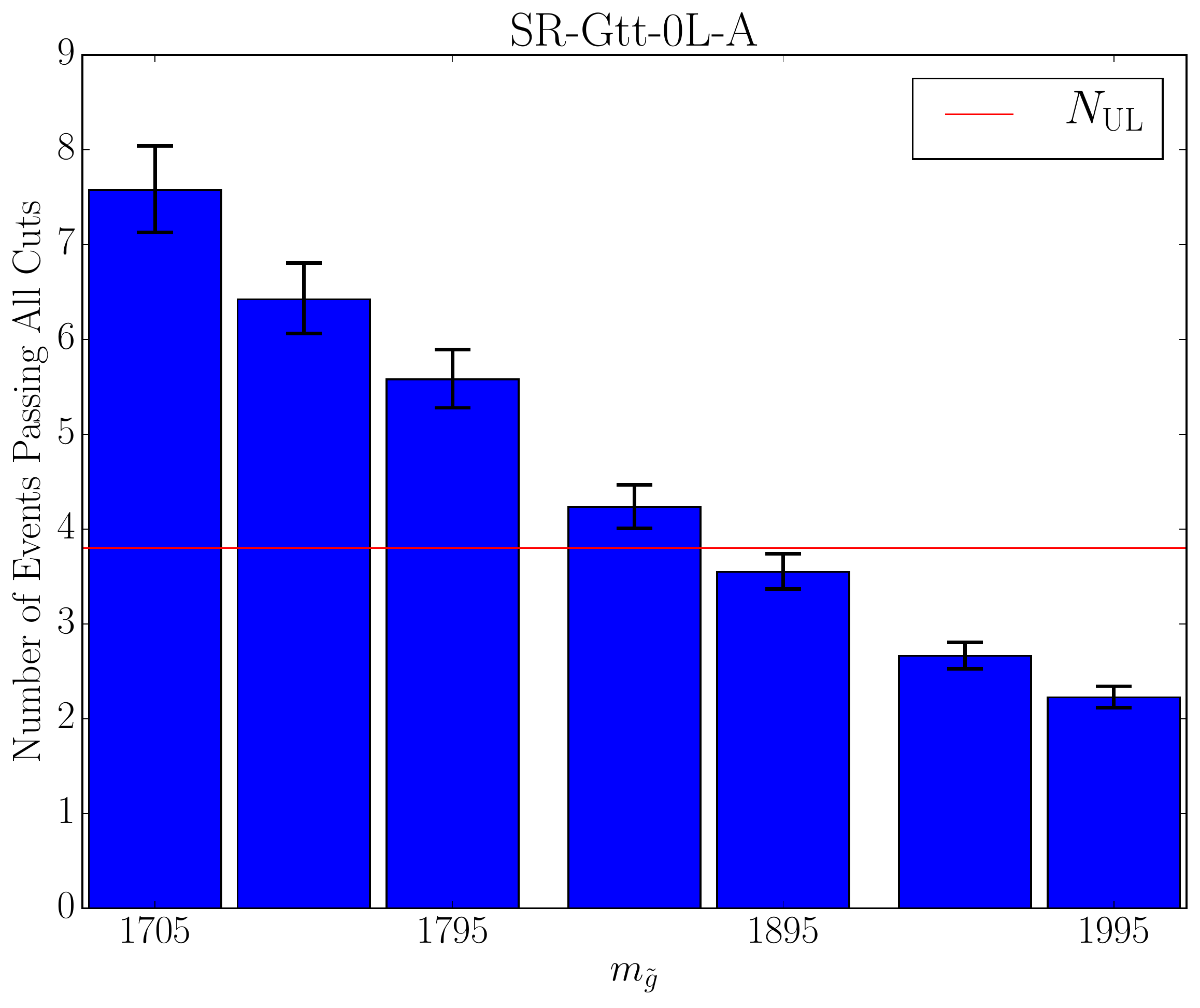}
    \caption{
    Number of events passing all cuts for our model in the 0-lepton with large mass splitting signal region of the ATLAS analysis~\cite{ATLAS:2016uzr}.  The scalar mass of all the points in this plot is $m_{16}=20\,\text{TeV}$.  The gluino mass bound obtained from this plot, $M_{\tilde g}\sim1.875\,\text{TeV}$, is the same as that from the validation plot.  Hence, we conclude that the gluino mass bound of our model is the same as that of the simplified model.
    }
    \label{fig:gtt0la}
  \end{center}
\end{figure}

On the other hand, Fig.~\ref{fig:gtt0la} is the plot for our model.  The scalar mass for all points in this plot is $m_{16}=20\,\text{TeV}$.  We have checked that points with $m_{16}=25\,\text{TeV}$ produce the same gluino mass bound.  Since the gluino mass bound of our model is very similar to that of the simplified model.   We conclude that the current gluino mass bound of our model $M_{\tilde g}\gtrsim1.9\,\text{TeV}$.

With this gluino mass bound, our model is unable to fit the excess CMS found in the two opposite-sign same-flavor leptons analysis~\cite{CMS:2016zwk}.  Hence, we predict that this excess is a statistical fluctuation.

\section{Predictions and Discovery Prospects}
\label{sec:ch5.predictions}
From the analysis in the previous section, the current gluino mass bound of our model is $M_{\tilde g}>1.9\,\text{TeV}$.  Even with this mass bound, Fig.~\ref{fig:contour} shows that a wide range of parameter space still is $<1.2\sigma$.  Hence, our model is not ruled out by low energy data and current LHC bounds.  Since our $\chi^2$ analysis is well below $2\sigma$ even for a gluino as heavy as $2.7\,\text{TeV}$, see Fig.~\ref{fig:contour}, this model will, unfortunately, not be ruled out, even if the gluino is not found in this run of the LHC.

The SUSY mass spectrum for two $m_{16}=20\,\text{TeV}$ points and two $m_{16}=25\,\text{TeV}$ points are given in Tab.~\ref{tab:spectrum}.  These points have $M_{\tilde g}=2.0\,\text{TeV}$ or $M_{\tilde g}=2.6\,\text{TeV}$.  The lightest scalar mass of our model, $m_{\tilde t_1}$ ranges from $3-5\,\text{TeV}$ while the first two families scalar masses are either around $20\,\text{TeV}$ or $25\,\text{TeV}$ depending on the value of $m_{16}$ at the GUT scale.  The scalars of the first two families are decoupled from the low energy theory while the third family scalars are not.  Hence, SUSY is not completely decoupled from the SM.  The CP-odd Higgs, $A$, the heavy Higgs, $H^0$, and the charged Higgs, $H^\pm$, all have masses around $5-6\,\text{TeV}$ showing that we are in the decoupling limit where the light Higgs behaves like a SM Higgs.  Tab.~\ref{tab:spectrum} also shows our prediction for the electron electric dipole moment, $\text{edm}_e$, the branching ratio $\text{BR}(\mu\to e\gamma)$ and the CP violating phase in the neutrino sector, $\sin\delta$.  These values are consistent with current experiment bounds.  Note, however, that these predictions differ significantly from our previous results.  In particular,  the electric dipole moment of the electron and the branching ratio, $BR(\mu \rightarrow e \gamma)$, are significantly smaller than before.  In addition, the CP violating angle in the lepton sector is now of order 90$^\circ$ for $m_{16} = 25$ TeV.

\begin{table}[!htbp]
  \begin{center}
    \renewcommand{\arraystretch}{1.0}
    \begin{tabular}{|l|r|r||r|r|}
      \hline
      $m_{16}/\text{TeV}$ & 20 & 25 & 20 & 25 \\
      $M_{\tilde g}/\text{TeV}$ & 2.00 & 2.00 & 2.60 & 2.60 \\
      $\chi^2/\text{dof}$ & 1.14 & 1.16 & 1.18 & 1.17 \\\hline
      $m_{\tilde t_1}/\text{TeV}$ & 3.68 & 4.70 & 3.70 & 4.65 \\
      $m_{\tilde t_2}/\text{TeV}$ & 4.38 & 5.52 & 4.43 & 5.49 \\
      $m_{\tilde b_1}/\text{TeV}$ & 4.17 & 5.32 & 4.17 & 5.23 \\
      $m_{\tilde b_2}/\text{TeV}$ & 4.32 & 5.47 & 4.36 & 5.43 \\
      $m_{\tilde\tau_1}/\text{TeV}$ & 7.47 & 9.30 & 7.52 & 9.27 \\
      $m_{\tilde\tau_2}/\text{TeV}$ & 12.2 & 15.2 & 12.2 & 15.2 \\
      $m_{\tilde\chi_1^0}/\text{GeV}$ & 352 & 352 & 474 & 474 \\
      $m_{\tilde\chi_2^0}/\text{GeV}$ & 586 & 636 & 650 & 665 \\
      $m_{\tilde\chi_1^+}/\text{GeV}$ & 585 & 636 & 646 & 661 \\
      $m_{\tilde\chi_2^+}/\text{GeV}$ & 710 & 751 & 911 & 914 \\
      $(M_A\approx M_{H^0}\approx M_{H^\pm})/\text{TeV}$ & 5.18 & 6.39 & 5.39 & 6.67 \\\hline
      $\text{edm}_e/10^{-32}\,\text{e cm}$ & -3.46 & -1.77 & -4.47 & -2.28 \\
      $\text{BR}(\mu\to e\gamma)/10^{-17}$ & 2.08 & 0.922 & 1.84 & 0.869 \\
      $\sin\delta$ & 0.759 & 0.935 & 0.644 & 0.993 \\\hline
    \end{tabular}
  \end{center}
  \caption{
  SUSY mass specturm from two $m_{16}=20\,\text{TeV}$ points and two $m_{16}=25\,\text{TeV}$ points of our model.
  The points have $M_{\tilde g}=2.0\,\text{TeV}$ and $M_{\tilde g}=2.6\,\text{TeV}$ respectively.
  In addition, the prediction of the electron dipole moment, the branching ratio of $\mu\to e\gamma$ and the neutrino CP violating phase are also presented in this table.
  }
  \label{tab:spectrum}
\end{table}

\section{Conclusion}
\label{sec:ch6.conclusion}
In this paper, we modify the Yukawa sector of a SUSY GUT with SO(10) or PS gauge symmetry that some of the present authors have studied in the past.  This paper aims to improve the fits to low energy observables, such as $\sin2\beta$, $m_u$ and $m_d$.  By shifting the phase from one Yukawa texture to another, we are able to improve the fit to $\sin2\beta$.  On the other hand, to fit $m_u$ and $m_d$, we choose PS gauge symmetry, due to higher flexibility than SO(10), and introduce two real parameters to the 11 and 12/21 entries of the Yukawa matrices.  This increases the number of parameters of our model to 26 parameters, see Tab.~\ref{tab:input}.

By fitting to 51 low energy observables, see Tab.~\ref{tab:obs}, our global $\chi^2$ analysis has 27 dof.  The modification to the Yukawa sector, see Sec.~\ref{sec:ch2.model}, improves the best fit from $\chi^2/\text{dof}=1.90$~\cite{Bryant:2016sjj} to $\chi^2/\text{dof}=1.12$, see Fig.~\ref{fig:contour}.  Even for gluinos as heavy as $M_{\tilde g}=2.7\,\text{TeV}$, our analysis shows that $\chi^2/\text{dof}\approx1.15$.  Thus, our model will not be ruled out even if gluinos are not found during this LHC run.  On the bright side, our model indicates that low energy SUSY is still a viable model and LHC might hopefully find gluinos in the near future.

In addition to the global $\chi^2$ analysis, we also reinterpreted ATLAS and CMS analyses in signal regions with high jet multiplicities and large missing transverse momentum.  Since gluinos of our model do not decay via a single decay channel, see Tab.~\ref{tab:gluino_br}, the gluino mass bound of our model might be different from that of a simplified model.  Gluinos of our model decay predominantly via $tb\tilde\chi_{1,2}^+$, $t\bar{t}\tilde\chi_{1,2,3,4}^0$ and $b\bar{b}\tilde\chi_{1,2,3,4}^0$.  However, we found that the gluino mass bound of our model is very similar to that of a simplified model where $M_{\tilde g}\sim1.9\,\text{TeV}$.  The most constraining signal region that we found is also the same as that of the simplified model, which is from the ATLAS analysis in the signal region with 0-lepton, large jet multiplicities and large missing transverse momentum (Gtt-0L-A) of \uppercase{atlas-conf-2016-052}~\cite{ATLAS:2016uzr}.

Previous analysis by Bryant et.~al.~shows that this model can be extended to fit inflation observables measured by BICEP2/Keck and Planck joint collaboration via a subcritical hybrid inflation~\cite{Bryant:2016tzg}.  Further studies of the consequences of this model for the early universe are warranted.

Since SUSY particles have not been observed at the LHC and natural SUSY models prefer light superpartners, one might think that low energy SUSY models are no longer attractive on the grounds of naturalness.  However, previous analysis by Poh et.~al.~showed that the fine-tuning of this model can be of order of 1 part in 500~\cite{Poh:2015wta}, assuming that some soft SUSY breaking boundary conditions defined at $M_{GUT}$ can be obtained from a more fundamental theory.  Thus, although this model is not the most natural model, it is much more natural than the SM.  In addition, by construction, this model uses small GUT representations, thus the model has the potential for a UV completion to a higher dimensional string theory.

\section*{Acknowledgments}
We are indebted to Radovan Derm\'i\v{s}ek for his program and his valuable inputs in using it.
Z.P.~and S.R.~received partial support for this work from DOE/DE-SC0011726.
We are also grateful to B. Charles Bryant, Christopher Hill and Weifeng Ji for discussions.

\appendix
\section{Benchmark Point}
\begin{table}[!htbp]\footnotesize
  \renewcommand{\arraystretch}{1.0}
  \begin{center}
    \begin{tabular}{|c|l|r|c|c|l|r|}
      \multicolumn{7}{l}{\normalsize Benchmark point with $m_{16} = 20.0\,\text{TeV},M_{\tilde g}=2.00\,\text{TeV}$}
      \\\cline{1-3}\cline{5-7}
      Sector & Input Param. & Best Fit & &
      Sector & Input Param. & Best Fit
      \\\cline{1-3}\cline{5-7}\cline{1-3}\cline{5-7}
      \multirow{3}{*}{Gauge}
        & $1/\alpha_G$ & 26.0 & &
      \multirow{13}{*}{Yukawa Textures}
        & $\lambda$ & 0.617 \\
        & $M_G/10^{16}\,\text{GeV}$ & 2.25 & &
        & $\lambda\epsilon$ & 0.0326 \\
        & $\epsilon_3/\%$ & -1.68 & &
        & $\lambda\tilde\epsilon$ & 0.0100 \\\cline{1-3}
      \multirow{5}{*}{SUSY (GUT scale)}
        & $m_{16}/\text{TeV}$ & 20.0 & &
        & $\lambda\epsilon'$ & -0.00300 \\
        & $m_{1/2}/\text{GeV}$ & 660 & &
        & $\lambda\xi$ & 0.00201 \\
        & $A_0/\text{TeV}$ & -40.6 & &
        & $\alpha$ & 0.138 \\
        & $(m_{H_d}/m_{16})^2$ & 1.98 & &
        & $\beta$ & 0.0277 \\
        & $(m_{H_u}/m_{16})^2$ & 1.61 & &
        & $\theta'/10^{-5}$ & 5.03 \\\cline{1-3}
      \multirow{3}{*}{Neutrino}
        & $M_{R_1}/10^9\,\text{GeV}$ & 4.62 & &
        & $\tilde\theta/10^{-5}$ & 2.92 \\
        & $M_{R_2}/10^{11}\,\text{GeV}$ & 8.32 & &
        & $\phi_{\epsilon'}/\text{rad}$ & -0.277 \\
        & $M_{R_3}/10^{13}\,\text{GeV}$ & 4.71 & &
        & $\phi_\xi/\text{rad}$ & 3.41 \\\cline{1-3}
      \multirow{2}{*}{SUSY (EW Scale)}
        & $\tan\beta$ & 50.4 & &
        & $\phi_\alpha/\text{rad}$ & 0.963 \\
        & $\mu/\text{GeV}$ & 630 & &
        & $\phi_\beta/\text{rad}$ & -1.26 \\\cline{1-3}\cline{5-7}
    \end{tabular}
  \end{center}
\end{table}

\begin{table}[!htbp]\footnotesize
  \begin{center}
    \scalebox{0.8}{
      \renewcommand{\arraystretch}{1.0}
      \begin{tabular}{|l|r|r|r|r|}
        \multicolumn{5}{l}{\normalsize Benchmark point with $m_{16} = 20.0\,\text{TeV},M_{\tilde g}=2.00\,\text{TeV}$}
        \\\hline
        Observable & \multicolumn{1}{c}{Fit} & \multicolumn{1}{|c}{Exp.} & \multicolumn{1}{|c}{Pull} & \multicolumn{1}{|c|}{$\sigma$} \\
        \hline\hline
        $M_Z/\text{GeV}$ & 91.1876 & 91.1876 & 0.0000 & 0.4514 \\
        $M_W/\text{GeV}$ & 80.4734 & 80.3850 & 0.2238 & 0.3949 \\
        $1/\alpha_\text{em}$ & 137.3435 & 137.0360 & 0.4478 & 0.6867 \\
        $G_\mu/10^{-5}\,\text{GeV}^{-2}$ & 1.1761 & 1.1664 & 0.8264 & 0.0118 \\
        $\alpha_3(M_Z)$ & 0.1177 & 0.1181 & 0.4791 & 0.0008 \\
        \hline
        $M_t/\text{GeV}$ & 174.0978 & 173.2100 & 0.4161 & 2.1338 \\
        $m_b(m_b)/\text{GeV}$ & 4.3264 & 4.1850 & 1.0388 & 0.1362 \\
        $m_\tau/\text{Mev}$ & 1776.0100 & 1776.8600 & 0.0428 & 19.8568 \\
        \hline
        $(M_b-M_c)/\text{GeV}$ & 3.3028 & 3.4500 & 0.4098 & 0.3592 \\
        $m_c(m_c)/\text{GeV}$ & 1.2685 & 1.2700 & 0.0442 & 0.0332 \\
        $m_s(2\,\text{GeV})/\text{Mev}$ & 97.7602 & 98.0000 & 0.0393 & 6.0987 \\
        $m_s/m_d(2\,\text{GeV})$ & 18.5692 & 19.5000 & 0.3843 & 2.0519 \\
        $Q$ & 21.5785 & 23.0000 & 0.6256 & 2.2725 \\
        $m_u(2\,\text{GeV})/\text{MeV}$ & 2.6880 & 2.3000 & 0.7758 & 0.5002 \\
        $m_d(2\,\text{GeV})/\text{MeV}$ & 5.2646 & 4.7500 & 1.1417 & 0.4508 \\
        $M_\mu/\text{MeV}$ & 105.2131 & 105.6584 & 0.2053 & 2.1690 \\
        $M_e/\text{MeV}$ & 0.5108 & 0.5110 & 0.0278 & 0.0057 \\
        \hline
        $|V_{ud}|$ & 0.9745 & 0.9742 & 0.0622 & 0.0049 \\
        $|V_{us}|$ & 0.2245 & 0.2248 & 0.2615 & 0.0013 \\
        $|V_{ub}|/10^{-3}$ & 3.9904 & 4.1300 & 0.2305 & 0.6056 \\
        $|V_{cd}|$ & 0.2244 & 0.2200 & 0.8509 & 0.0051 \\
        $|V_{cs}|$ & 0.9735 & 0.9950 & 1.2853 & 0.0167 \\
        $|V_{cb}|/10^{-3}$ & 44.1574 & 40.7500 & 1.4038 & 2.4272 \\
        $|V_{td}|/10^{-3}$ & 7.9898 & 8.2000 & 0.3378 & 0.6222 \\
        $|V_{ts}|/10^{-3}$ & 43.6115 & 40.0000 & 1.2691 & 2.8458 \\
        $|V_{tb}|$ & 0.9990 & 1.0090 & 0.3179 & 0.0314 \\
        $\sin2\beta$ & 0.6922 & 0.6910 & 0.0672 & 0.0173 \\
        $\epsilon_K/10^{-3}$ & 2.0225 & 2.2330 & 1.0379 & 0.2028 \\
        \hline
        $\Delta M_{B_s}/\Delta M_{B_d}$ & 43.7269 & 34.8479 & 1.0037 & 8.8463 \\
        $\Delta M_{B_d}/10^{-10}\,\text{MeV}$ & 2.9005 & 3.3540 & 0.7802 & 0.5812 \\
        \hline
        $m^2_{21}/10^{-5}\,\text{eV}^2$ &  7.3484 & 7.3750 & 0.0658 & 0.4044 \\
        $m^2_{31}/10^{-3}\,\text{eV}^2$ &  2.5096 & 2.5000 & 0.0726 & 0.1323 \\
        $\sin^2\theta_{12}$ & 0.2960 & 0.2975 & 0.0915 & 0.0166 \\
        $\sin^2\theta_{23}$ & 0.4419 & 0.4435 & 0.0599 & 0.0266 \\
        $\sin^2\theta_{13}$ & 0.0217 & 0.0215 & 0.1493 & 0.0010 \\
        \hline
        $M_h/\text{GeV}$ & 122.7975 & 125.0900 & 0.4854 & 4.7225 \\
        \hline
        $BR(b\to s\gamma)/10^{-6}$ & 299.9500 & 332.0000 & 0.2243 & 142.9017 \\
        $BR(B_s\to\mu^+\mu^-)/10^{-9}$ & 5.1836 & 2.9500 & 1.6808 & 1.3289 \\
        $BR(B_d\to\mu^+\mu^-)/10^{-9}$ & 0.1223 & 0.4000 & 1.8234 & 0.1523 \\
        $BR(B\to\tau\nu)/10^{-6}$ & 96.4950 & 106.0000 & 0.1822 & 52.1761 \\
        \hline
        $BR(B\to K^*\mu^+\mu^-)_{1\leq q^2\leq6\,\text{GeV}^2}/10^{-7}$ & 0.5456 & 0.3400 & 0.3567 & 0.5765 \\
        $BR(B\to K^*\mu^+\mu^-)_{14.18\leq q^2\leq16\,\text{GeV}^2}/10^{-7}$ & 0.7904 & 0.5600 & 0.1531 & 1.5055 \\
        $q_0^2(A_\text{FB}(B\to K^*\mu^+\mu^-))/\text{GeV}^2$ & 3.8492 & 4.9000 & 0.7921 & 1.3265 \\
        $F_L(B\to K^*\mu^+\mu^-)_{1\leq q^2\leq6\,\text{GeV}^2}$ & 0.7522 & 0.6500 & 0.2917 & 0.3503 \\
        $F_L(B\to K^*\mu^+\mu^-)_{14.18\leq q^2\leq16\,\text{GeV}^2}$ & 0.3514 & 0.3300 & 0.0725 & 0.2952 \\
        $P_2(B\to K^*\mu^+\mu^-)_{1\leq q^2\leq6\,\text{GeV}^2}$ & 0.0679 & 0.3300 & 1.4536 & 0.1803 \\
        $P_2(B\to K^*\mu^+\mu^-)_{14.18\leq q^2\leq16\,\text{GeV}^2}$ & -0.4333 & -0.5000 & 0.3381 & 0.1973 \\
        $P_4'(B\to K^*\mu^+\mu^-)_{1\leq q^2\leq6\,\text{GeV}^2}$ & 0.5788 & 0.5800 & 0.0029 & 0.4007 \\
        $P_4'(B\to K^*\mu^+\mu^-)_{14.18\leq q^2\leq16\,\text{GeV}^2}$ & 1.2177 & -0.1800 & 1.7055 & 0.8195 \\
        $P_5'(B\to K^*\mu^+\mu^-)_{1\leq q^2\leq6\,\text{GeV}^2}$ & -0.3221 & 0.2100 & 2.0721 & 0.2568 \\
        $P_5'(B\to K^*\mu^+\mu^-)_{14.18\leq q^2\leq16\,\text{GeV}^2}$ & -0.7119 & -0.7900 & 0.1545 & 0.5053 \\
        \hline
        \multicolumn{3}{|l|}{Total $\chi^2$} & 30.9061 & \\
        \hline
      \end{tabular}
    }
  \end{center}
\end{table}

\clearpage
\newpage
\bibliographystyle{utphys}
\bibliography{bibliography}

\end{document}